\documentclass[aps,prd,twocolumn,showpacs]{revtex4}
\usepackage{graphicx}
\usepackage{latexsym}
\usepackage{amsmath}
\usepackage{amsfonts}
\usepackage{amssymb}

\newcommand{\be}{\begin{equation}}
\newcommand{\ee}{\end{equation}}
\newcommand{\ben}{\begin{eqnarray}}
\newcommand{\een}{\end{eqnarray}}
\newcommand{\sech}{\rm sech}

\begin{document}


\title{A domain wall model for spectral reflectance of plant leaves}
\author{Francisco A. Brito$^a$ and Morgana L.F. Freire$^b$ }
\address{$^a$Departamento de F\'\i sica, Universidade Federal de Campina Grande,\\
Caixa Postal 10071, 58109-970 Campina Grande, Para\'\i ba, Brazil\\
$^b$Departamento de F\'\i sica, Universidade Estadual da Para\'\i
ba, Campina Grande, 58100-001 Campina Grande, Para\'\i ba, Brazil}

\date{\today}

\begin{abstract}\noindent
We model a plant leaf by using two-dimensional domain walls with
internal structures. Such domain walls can be found as soliton
solutions in field theory describing magnetic materials. The
radiation scattered by such domain walls behaves quite similar to
the spectral reflectance of plant leaves. The model nicely simulates
the spectral reflectance of a plant leaf as a function of the
wavelength.
\end{abstract}
 \maketitle

\newpage

\begin{figure}[ht]
\centerline{\includegraphics[{angle=90,height=5.0cm,angle=0,width=9.0cm}]
{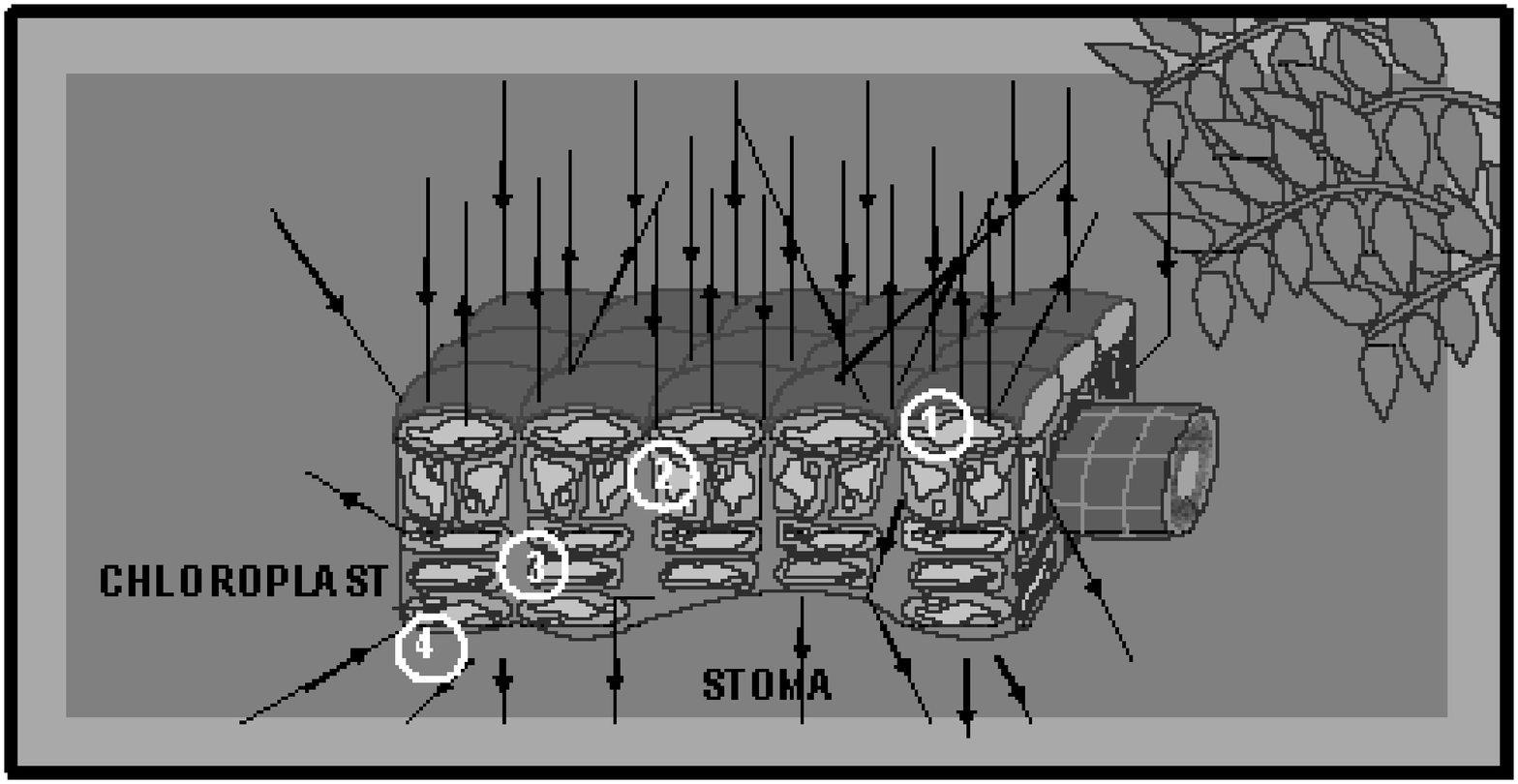}} \caption{Transversal section of a plant leaf with possible
paths of radiation rays. (1) upper epidermis, (2) palisade
mesophyll, (3) spongy mesophyll and (4) lower
epidermis.}\label{fig1}
\end{figure}

{\it I. Introduction.} The study of the spectral behavior of
vegetables is often used to investigate the characteristics of the
electromagnetic radiation reflected by plant leaves, a plant as
whole or a collection of plants. We should emphasize that the study
of the spectral behavior in general means the study of spectral
reflectance, spectral transmittance and spectral absorbance.
Concerning the interaction of the solar energy with vegetables, the
plant leaf plays a special role because it basically realizes the
photosynthesis that is responsible by the formation of carbon
compounds. The whole configuration of a plant leaf including form,
position, structures, etc, adapts to receive in a most efficient way
the solar beams, air and water which are necessary for realizing
photosynthesis. Thus the knowledge of the properties of a plant leaf
is clearly fundamental, especially in the study of the reflectance
of a plant or even a culture of plants. Such predominance of the
plant leaf is so important that the area of the other part of the
plant in contact with the solar radiation is completely neglected
\cite{1,1a}. It is obvious, however, that the data obtained from a
unique plant leaf cannot be used directly, without modifications, to
a culture of plants. This is because there exist qualitative and
quantitative differences between the aspects of unique plant leaf
and a culture of plants in a field. The structures of the cells that
constitute the three tissues of the leaves vary with the specie and
ambient conditions. The compounds of the leaf that are of
considerable importance in the study of interaction of a leaf with
the radiation are: cellulose that are found in the cellular walls,
solutes like ions and molecules, intercellular spaces and pigment
inside chloroplasts such as carotene, xanthophyll and chlorophyll.
The spectral behavior of a plant leaf is a function of its
compounds, morphology and internal structure. Since the
characteristics of the plant leaves are genetically controlled there
exist differences in the spectral behavior among groups that are
genetically distinct. The Fig.~\ref{fig1} shows possible paths of
the radiation or incident energy over a plant leaf. A small amount
of radiation is reflected from the cells around the surface of the
plant leaf; the major part of the radiation is transmitted to the
spongy mesophyll, where the cellular walls for incident angles
sufficiently large reflect the incident radiation rays \cite{2}.
These multiple reflections are essentially a random process where
the rays change their directions inside the plant leaf. Because the
large number of cellular walls, some rays go back toward the source
of radiation, while others are transmitted trough the plant leaf.
The thickness of the plant leaf clearly affects the transmittance of
the radiation, being larger than the reflectance for thin leaves.
The opposite happens for thick leaves. The spectral response of a
target, as a plant leaf considered in the present work, deal with
the graphic representation of the reflectance in narrow and adjacent
bands of wavelengths. This representation gives us the effect of the
interaction of incident radiation and the target under
investigation. Thus, the amplitude variations in spectral response
give us clues about the spectral properties of the objects
\cite{3,4,7,8}. The curve of reflectance of a plant leaf can be
divided into ultraviolet region, visible region (400 - 700 nm), near
infrared (700 - 1300 nm), and mid infrared (1300 - 2600 nm).
However, the ultraviolet region usually is not considered because
the most part of this radiation is scattered away and the vegetables
does not make use of this radiation. The mid infrared is not
considered here because of limitations on the sensor equipment in
this region. The curve of reflectance obtained experimentally is
shown in Fig. 2.

In this work, we model the plant leaf by a domain wall with internal
structure \cite{9,10,11,12,13,14}. In the following section we show
with details how this can be possible. The model is mainly based on
the reflection probability that light neutral particles can suffer
as they collide with such domain walls --- in Ref.~\cite{9} this
issue was initiated in another context. In our model the light
neutral particles play the role of the electromagnetic radiation and
the domain wall with internal structure plays the role of a plant
leaf.

The internal structures are small domain walls that we identify with
the spaces among the spongy mesophyll. Domain walls are soliton
solutions that appear in systems that exhibit spontaneous symmetry
breaking, such as the Higgs mechanism applied in Particle Physics
and the Ginzburg-Landau theory applied in Superconductivity. These
systems usually present non-linearity into equations of motion. Such
non-linearity is responsible for the formation of soliton solutions
whose energy density is localized around a region of the space, and
remains stable all times.

\begin{figure}[ht]
\centerline{\includegraphics[{angle=90,height=7.0cm,angle=0,width=8.0cm}]
{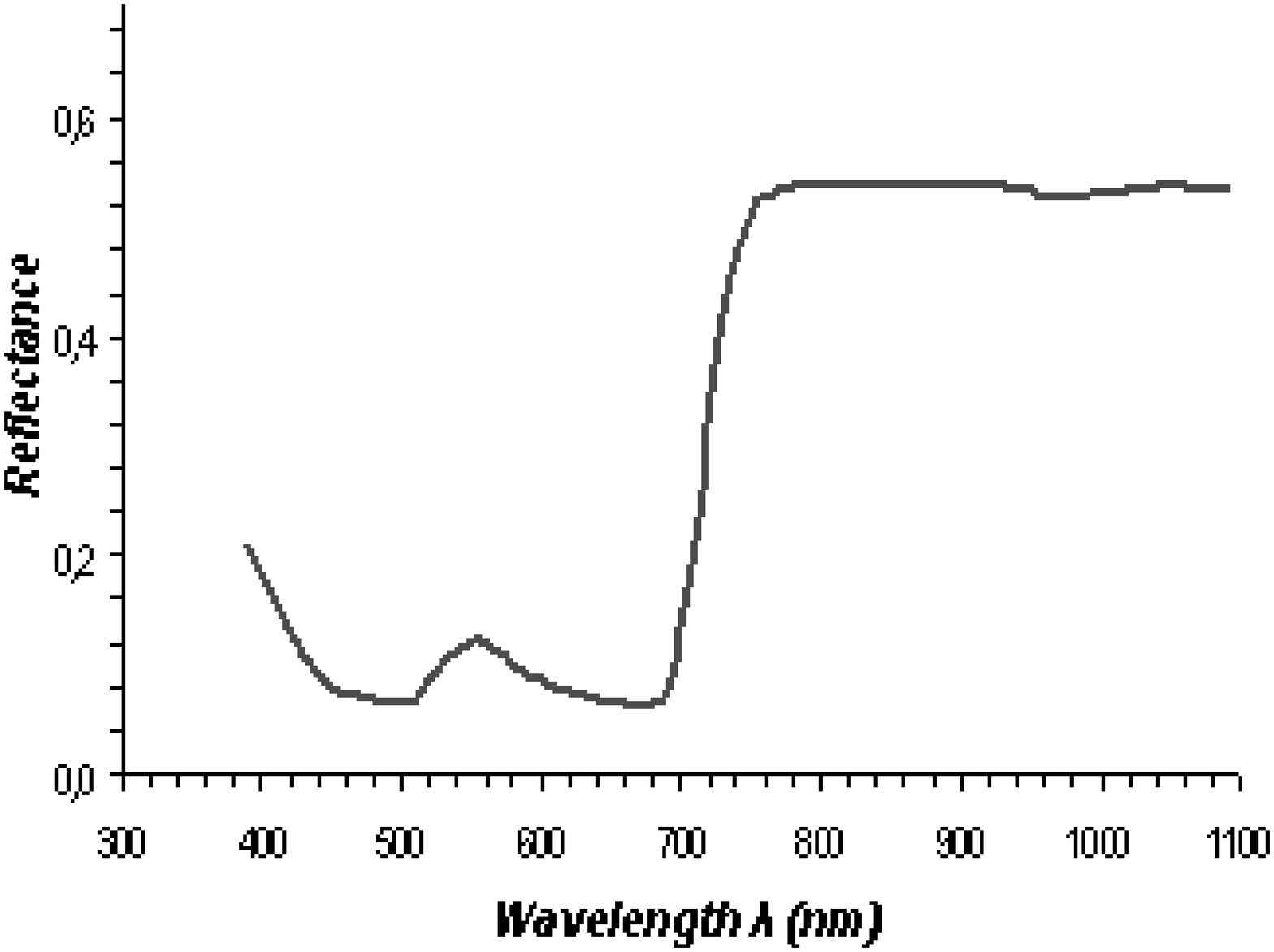}} \caption{The spectral reflectance of a peanut leaf in the
visible and infrared regions obtained experimentally.}\label{fig2}
\end{figure}

{\it II. The domain wall model.} The dynamics of a system with two
coupled real scalar fields $\phi$  and $\chi$  is described by a
field theory with the Lagrangian \ben\label{model}{\cal L}=
\frac{1}{2}\partial_\mu\phi\partial^\mu\phi+
\frac{1}{2}\partial_\mu\chi\partial^\mu\chi-V(\phi,\chi).\een Where
the potential $V(\phi,\chi)$ belongs to a class of soliton models
easily integrable \cite{9,10,11,12,13,14} given by \ben\label{Vw}
V=\frac{1}{2}\left(\frac{\partial W}{\partial\phi}\right)^2+
\frac{1}{2}\left(\frac{\partial W}{\partial\phi}\right)^2,\een where
$W$ is known as ``superpotential'' that for our purposes here it is
sufficient to make the following choice
\ben\label{W}W=\nu\left(\frac{\phi^3}{3}-a^2\phi\right)+\mu\phi\chi^2
,\een such that $V(\phi,\chi)$ can be written as \ben\label{pot}
V(\phi,\chi)&=&\frac{1}{2}\nu^2(\phi^2-a^2)^2+(2\mu^2+\nu\mu)\phi^2\chi^2
-\nu\mu a^2\chi^2\nonumber\\&+&\frac{1}{2}\mu^2\chi^4.\een The
scalar field $\phi$ describes the domain wall that we identify with
a plant leaf. The scalar field $\chi$ describes the internal
structures that appear as domain walls, which can form disk like
structures \cite{14}. These {\it domain walls} will be identified
later with {\it cellular walls} inside the plant leaf. The meaning
of the parameters $\nu$, $a$  and $\mu$ will become clear later. The
field equation of motion are given by
 \ben\label{eom1}\square\phi+\frac{\partial V}{\partial\phi}=0,\een
and \ben\label{eom2}\square\chi+\frac{\partial
V}{\partial\chi}=0.\een  Now we make use of perturbation theory, up
to first order, around the soliton solutions $\bar{\phi}$ and
$\bar{\chi}$  into the equations of motion (\ref{eom1}) and
(\ref{eom2}), as $\chi=\bar{\chi}+\xi$ and $\phi=\bar{\phi}$, where
$\xi$ represents the relevant fluctuations for the internal
structures. Considering such fluctuations having the form
$\xi=\xi(z)e^{-i(\omega t-k_x x - k_y y)}$ the relevant equation for
the fluctuations reads
\ben\label{flut}\partial_z^2\xi(z)+[-\omega^2+k_x^2+k_y^2]\xi(z)
+\frac{\partial^2\overline{V}}{\partial \chi^2}\xi(z)=0, \een which
is a Schroedinger like equation. The bar means the function is
evaluated at the original soliton solutions. By considering a domain
wall approximately plane and static it suffices to work only with a
spatial dimension such that the fields are functions as
$\phi\equiv\phi(z)$ and $\chi\equiv\chi(z)$. The equations of motion
have the following well-known solutions \cite{11,13}
\ben\label{typeI} && \phi=-a\tanh{(\nu a z)},\:\: \chi=0,
\\
\label{typeII} && \phi=-a\tanh{(2\mu a z)},\:\: \chi\!=\!\pm
a\sqrt{\frac{\nu}{\mu}\!-\!2}\,{\sech}\,(2\mu a z). \een Where $z$
is a coordinate that is transversal to the domain wall. The type I
solution (\ref{typeI}) represents a domain wall $without$ internal
structure while the type II  solution (\ref{typeII}) represents a
domain wall $with$ internal structure. Both solutions have the same
energy (or rest mass) $M_W=\sigma_W A_W$  with surface density
\ben\label{tension1} \sigma_W=\frac{4}{3}\nu a^3,\een that can be
found by taking the difference of the superpotential evaluated at
the vacuum solutions $\bar{\phi}=\pm a$ and $\bar{\chi}=0$ , i.e.,
$\sigma_W=|\Delta W|$ - see references \cite{9,10,11,12,13,14}. In
the type II solution, as the field $\phi\to0$ in the interior of the
domain wall, the field $\chi$ develops a nonzero maximum value that
characterizes its localization into the domain wall. In such regime
the potential $V(\phi,\chi)$ given in (\ref{pot}) approaches
effectively the potential
\ben\label{pot_eff}V_{eff}=\frac{1}{2}\mu^2\left(\chi^2-
\frac{\nu}{\mu}a^2\right)^2.\een The effective Lagrangian that now
governs the dynamics inside the domain wall is given by
\ben\label{Leff} {\cal
L}_{eff}=\frac{1}{2}\partial_\mu\chi\partial^\mu\chi-V_{eff}(\chi).\een
Given the potential (\ref{pot_eff}), this theory is able to describe
internal structures such as other domain walls given by the
following solution \ben\chi=\pm
a\sqrt{\frac{\nu}{\mu}}\tanh{\left(\mu\sqrt{\frac{\nu}{\mu}}x_i\right)},
\een where $x_i$ is some tangential coordinate along the domain wall
that is going to be identified with a plant leaf. The structures
have energy (or rest mass) $M_S=\sigma_S A_S$  with surface density
given by \ben\label{tension2}\sigma_S=\frac{4}{3}\nu
a^3\sqrt{\frac{\nu}{\mu}}. \een This can be found by constructing
the ``effective superpotential''
$W_{eff}=\mu\left(\frac{\chi^3}{3}-\frac{\nu}{\mu}a^2\chi\right)$
that produces the effective potential given by (\ref{pot_eff}). As
in the previous case, the surface density is found by evaluating the
superpotential at vacuum solutions $\bar{\chi}\!=\!\pm
a\sqrt{\frac{\nu}{\mu}}$, such that $\sigma_S=|\Delta W_{eff}|$.

Now using the type II solution (\ref{typeII}) into equation
(\ref{flut}) we find \ben\label{eta_sch}
\xi''(z)+k_z^2\xi(z)-U(z)\xi(z)=0,\een where
$-k_z^2\!=\!-\omega^2+k_x^2+k_y^2+m_\chi^2$ is the $z$-component of
the particle momentum, $m_\chi^2\!=\!4\mu^2a^2$  is the squared mass
of the particles far from the domain wall,  and the Schroedinger
potential is given by \ben\label{Usch}
U(z)=m_\chi^2\left[4-\frac{\nu}{\mu}{\rm sech}^2{(2\mu
az)}\right].\een The reflection probability related to the potential
$U(z)$ is well-known \cite{15}
\ben\label{eq20}R\!=\!\frac{\cos^2{\beta}}{\sinh^2{\gamma}+\cos^2{\beta}},
\een  with $\beta\!=\!(\pi/2)\sqrt{17-4\nu/\mu}$ and $\gamma\!=\!\pi
k_z/2\mu a$. However, by noting that the rest mass of the domain
wall can be related to the rest mass of the internal structures as
$M_W\!=\!NM_s$, we can write \ben \label{eq21}\sigma_W
V_W\!=\!N\sigma_S V_S\to A_W\!=\!\sqrt{\frac{\nu}{\mu}}NA_S\to
\sqrt{\frac{\nu}{\mu}}\approx\frac{\lambda^2}{\lambda^2_0}.\een Here
we used the fact  $V\!=\!A\delta$, being $\delta$  the domain wall
thickness. Thus $\beta\!=\!(\pi/2)\sqrt{17-4\lambda^4/\lambda_0^4}$,
and since $k_z\!=\!2\pi/\lambda$ and $\delta\!=\!1/2\mu a$ we find
$\gamma\!=\!2\pi^2\delta/\lambda$. Where we have approximated the
area of the internal structures as the area of a disk like structure
$A_S\approx2\pi L^2$ and considered that the typical size of the
structures is related to a critical wavelength as $L\approx
\lambda_0$ . On the other hand, by considering $A_W\approx2\pi
N\lambda^2$ we define $A_W/N\approx 2\pi \lambda^2$ as the smallest
area of radius $\lambda$ covered by a radiation with wavelength
$\lambda$.

\begin{figure}[ht]
\centerline{\includegraphics[{angle=90,height=7.0cm,angle=0,width=8.0cm}]
{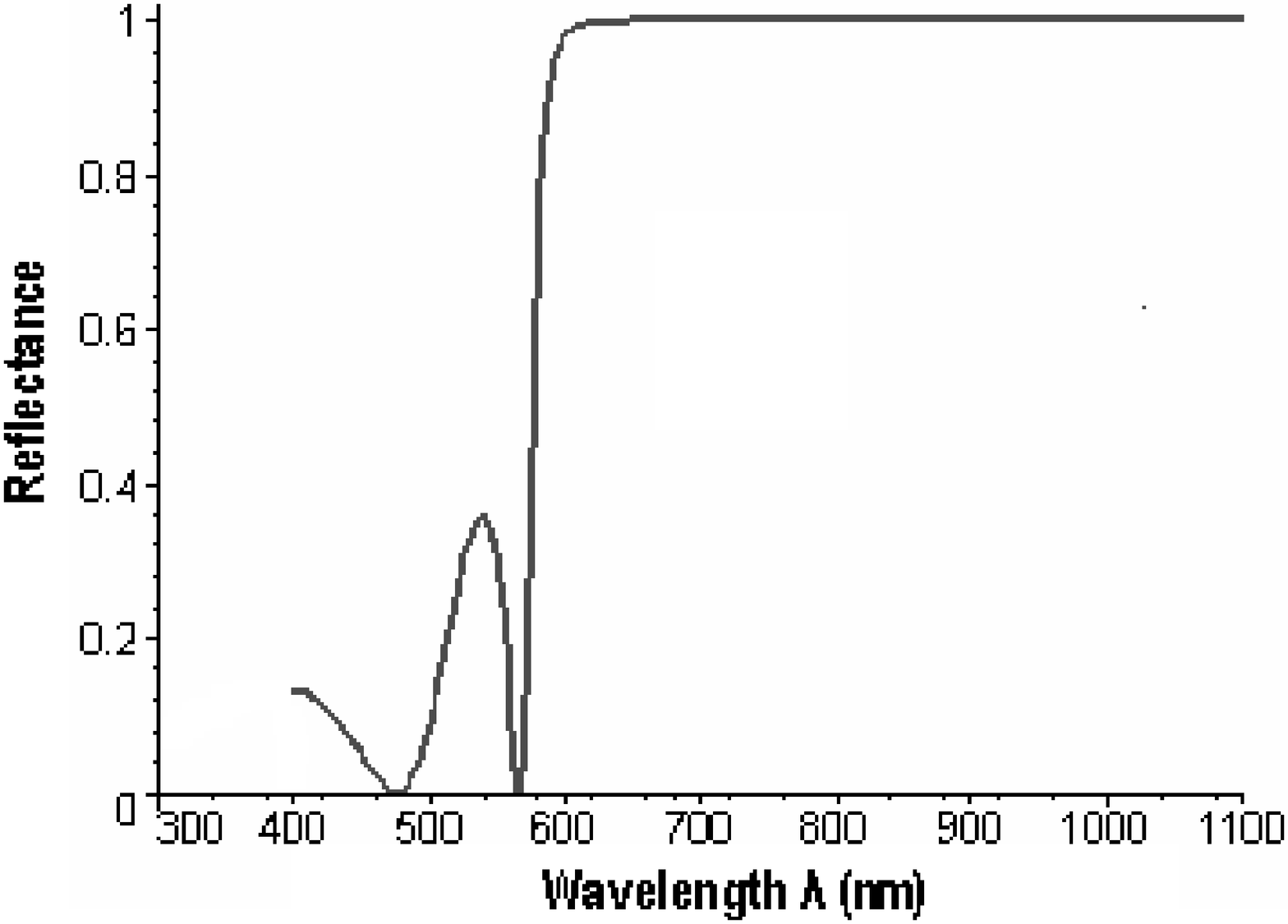}} \caption{The spectral reflectance of a plant leaf according
to the domain wall model, for $\lambda_0=400\,nm$  and
$\delta=30\,nm$.}\label{fig3}
\end{figure}

{\it III. Results and discussions.} The Figures \ref{fig2} and
\ref{fig3} show the reflectance curves obtained experimentally via
spectrum-radiometer \footnote{This concerns an equipment non-imager
that was linked to a spherical integrator LI-1800 via cable of
optical fiber. The light used is an artificial one that simulates
the sunlight. Such a configuration allowed the measure of the
spectral reflectance of a single plant leaf. The spectral resolution
of spectrum-radiometer is $2\,nm$.} LI-1800 from LI-COR and
theoretically via domain wall model, respectively. As we can see in
these figures, the model (Fig.~\ref{fig3}) nicely simulates the real
spectral behavior of a plant leaf (Fig.~\ref{fig2}). In the visible
region (400 a $700\,nm$) one observes two valley around $480\, nm$
and $680\, nm$ that evidences the absorption of the chlorophylls
``b'' and ``a'', respectively, and a peak around $555\, nm$ that
characterizes the color of a green plant leaf. In the near infrared
region ($700\, nm$ - $1300\, nm$) the reflectance is almost
constant. Let us now compare the results obtained experimentally
with the results obtained theoretically via domain wall model, by
entering with some known parameters of the model. We can use the
formula below to estimate the critical wavelength \ben\label{eq22}
\lambda_0\!=\!\sqrt{\frac{A_S}{2\pi}}.\een For structures (or
spaces) with typical dimensions about 4-12 $\mu m$ of width and
10-14 $\mu m$ of length, as the regions separating spongy
mesophylls, we find critical wavelength $\lambda_0\approx400\, nm$.
The parameter $\delta$ is associated with the domain wall thickness,
and can be regarded as an {\it effective thickness} of the plant
leaf, where effectively reflections may occur, i.e., the interior
region of the plant leaf. The Fig.~\ref{fig3} shows the theoretical
behavior of the spectral reflectance of a plant leaf according to
our model. Unlike the spectral reflectance obtained experimentally
(see Fig.~\ref{fig2}) the minimal reflectance is zero because only
internal effects are considered in the model. For the sake of
simplicity, we disregard reflections from the surface of the plant
leaf. Note that the spectral reflectance from internal structures of
plant leaves is well approximated by our domain wall model. Although
the second point of absorption around 565 $nm$ is a bit far from
observed value 680 $nm$, the first point of absorption at 480 $nm$
is in perfect agreement and the reflectance peaked around 540 $nm$
is pretty close to the observed value around 555 $nm$ that
characterizes the color of a green plant leaf.

{\it IV. Conclusions.} The model nicely simulates the spectral
reflectance of a plant leaf as a function of the wavelength, i.e.,
its spectral behavior. The domain wall model works well for
structures in the scale of micrometers ($\mu m$ ). This is also the
scale of spaces between cells ( $\approx$ cellular walls thickness),
that we can identify with the domain wall like structures in our
model discussed above. In order to simulate better the spectral
behavior of a plant leaf we expect to improve the model in a
separate work. Such improvements may be concerned with the number
and the type of domain walls like structures formed inside the
domain wall that simulate the plant leaf. This may improve the
description of several intercellular spaces giving the real
complexity of the internal structure inside a plant leaf. Finally we
should emphasize that our investigations concerning a domain wall
model to simulate spectral reflectance here applied to a plant leaf
can easily be extended to a class of planar systems with internal
structures.

\acknowledgments

The author F.A. Brito thanks CNPq for partial support.



\end{document}